\documentclass[11pt]{article}
\usepackage{amsmath,amsfonts,amssymb}
\usepackage{graphicx,color}
\usepackage{epsfig,epsf}
\usepackage[toc,page]{appendix}
\usepackage{bm}
\usepackage{dcolumn,subfig}
\usepackage{array}
\usepackage{cite}
\usepackage{color}

\usepackage{relsize}
\def\Babar{{\mbox{\slshape B\kern-0.1em{\smaller A}\kern-0.1em B\kern-0.1em{\smaller A\kern-0.2em R}}}}

\usepackage[bookmarks, breaklinks, colorlinks,urlcolor=black, citecolor=red, 
linkcolor=blue]{hyperref}

\usepackage{booktabs}
\usepackage{graphicx}
\graphicspath{ {images/} }
\usepackage{amssymb}

\usepackage{graphicx}
\usepackage{graphics}
\DeclareGraphicsExtensions{.eps}
\usepackage{float} 
\usepackage{mathrsfs}
\usepackage{amsmath}
\usepackage{slashed}
\usepackage{amsmath}
\usepackage{amsfonts}   
\usepackage{amssymb}
\usepackage{subfloat}

 \usepackage{color}

 \usepackage[normalem]{ulem}

 \definecolor{darkgreen}{cmyk}{1,0,1,0.2}
 
\def\com2#1{\textcolor{red}{\it{#1}}}

\def\be {\begin{equation}}
\def\ee {\end{equation}}
\def\bea {\begin{eqnarray}}
\def\eea {\end{eqnarray}}
\def\n {\nonumber}
\def\bra {\langle}
\def\ket {\rangle}

\begin{document}

\renewcommand*{\thefootnote}{\fnsymbol{footnote}}

\begin{center}
 {\Large\bf{Hierarchy problem and dimension-six effective operators}}\\[5mm]
{\bf Ambalika Biswas} $^{a}$\footnote{ani73biswas@gmail.com},
 {\bf Anirban Kundu} $^{b}$\footnote{anirban.kundu.cu@gmail.com},
and 
{\bf Poulami Mondal} $^{b}$\footnote{poulami.mondal1994@gmail.com}\\[3mm]
${}^a$ Department of Physics, Vivekananda College,\\
269, Diamond Harbour Road, Thakurpukur, Kolkata 700063, India\\
${}^b$ Department of Physics, University of Calcutta, \\
92 Acharya Prafulla Chandra Road, Kolkata 700009, India
 \\ 
 \today
 \end{center}


\begin{abstract}

Without any mechanism to protect its mass, the self-energy of the Higgs boson diverges quadratically, leading 
to the hierarchy or fine-tuning problem. One bottom-up solution is to postulate some yet-to-be-discovered 
symmetry which forces the sum of the quadratic divergences to be zero, or almost negligible; this is known
as the Veltman condition. Even if one assumes the existence of some new physics at a high scale, 
the fine-tuning problem is not eradicated, although it is softer than what it would have been with a Planck scale 
momentum cut-off. We study such divergences in an effective theory framework, and construct 
the Veltman condition with dimension-six operators. We show that there are two classes of diagrams, the
one-loop and the two-loop ones, that contribute to quadratic divergences, but the contribution of the 
latter is suppressed by a loop factor of $1/16\pi^2$. There are only six dimension-six operators that contribute to the 
one-loop category, and the Wilson coefficients of these operators 
play an important role towards softening the fine-tuning problem. We find the parameter space for the 
Wilson coefficients that satisfies the extended Veltman condition, and also discuss why one need not bother 
about the $d>6$ operators. The parameter space is consistent with the theoretical and experimental bounds 
of the Wilson coefficients, and should act as a guide to the model builders.  

\end{abstract}



\setcounter{footnote}{0}
\renewcommand*{\thefootnote}{\arabic{footnote}}

\section{Introduction}

The title of this paper may appear to be an oxymoron for two reasons. First, effective theories are known to be 
valid up to a certain energy scale, so why should one talk about the hierarchy problem, which essentially is a 
manifestation of extreme weakness of gravity, or the extremely high value of the Planck scale $\sim 10^{19}$ GeV? 
Second, any calculation of the scalar self-energy involves the evaluation of loop contributions to the 
self-energy, and how may one evaluate a loop in an effective theory with higher-dimensional operators? 

Both these problems can be easily surmounted (see, {\em e.g.}, Ref.\ \cite{Bar-Shalom:2014taa}).  
In a sense, it is subjective, depending on what level of fine-tuning one is comfortable with. If the cutoff scale 
of the theory be $\Lambda$ and the Higgs vacuum expectation value (VEV) be $v$, the typical fine-tuning 
is of the order of $v^2/\Lambda^2$. One can, however, be more quantitative. One may write the physical
Higgs mass, $m_h$, in terms of a bare mass term $m_{h,0}$ and higher-order self-energy corrections:
\be
m_h^2 = m_{h,0}^2 + \delta m_h^2\,,
\ee
where $\delta m_h^2$ is some function of the masses, $v$, and $\Lambda$. In this case, 
$m_h^2/|\delta m_h^2|$ may be taken as an approximate measure of fine-tuning. 
For $\Lambda=2$ TeV, just outside the reach of the Large Hadron Collider (LHC), 
this is about one or a few percent, not at all uncomfortable, but higher values of $\Lambda$ definitely brings back 
the fine-tuning problem, maybe in a softened way. Also, if one has a renormalisable theory below $\Lambda$,
loop calculations do not pose any problem, with the understanding of a momentum cut-off at $\Lambda$. 
Cut-off regularisation is not Lorentz invariant, but it is undoubtedly the best way to feel the badness of
a divergence.

Very briefly, the hierarchy of fine-tuning problem is why the Higgs mass is at the electroweak scale and not at the 
Planck scale, when it is not protected by any symmetry. If we use a cut-off regularisation, the Higgs self-energy 
diverges as $\Lambda^2$ while the fermion and gauge boson masses diverge only logarithmically. Thus, to get 
a Higgs mass of the order of $v$ from a quantum correction of the order of $\Lambda$, one needs a fine-tuning
between the bare mass term and the quantum corrections. 

In this paper, we will not talk about any possible ultraviolet complete (UVC) theory, like supersymmetry, that 
may solve the hierarchy problem. We will, rather, demand that perhaps due to some 
yet-to-be-discovered symmetry, the quadratically divergent contributions to the Higgs mass add up to zero, 
or a very small value. This is known as the Veltman condition (VC) \cite{Veltman:1980mj}.

Thus, if we confine ourselves to one-loop diagrams only, the renormalised Higgs mass squared is given by
\be
m_h^2 = m_{h,0}^2 + \delta m_h^2 = m_{h,0}^2 + \frac{1}{16\pi^2} \, f(g_i) \Lambda^2 + \cdots \,,
\ee
where the ubiquitous coefficient of $1/16\pi^2$ comes from 
the evaluation of the loop, and $f(g_i)$ is a function of relevant scalar, Yukawa, and 
gauge couplings. The VC demands that $f(g_i)$ should be zero, or extremely tiny, so that $m_{h,0}$ 
is not too much away from the electroweak scale. The logarithmically divergent as well as the finite terms 
coming from the loop diagrams have been neglected, and denoted by the trailing ellipses. 

One may argue that $f(g_i)$ need not be exactly zero; in fact, $f(g_i) \sim 16\pi^2 m_h^2/\Lambda^2$ 
should be perfectly acceptable. However, with all the masses known, the VC fails badly for the SM 
\cite{Grzadkowski:2009mj,Einhorn:1992um}. There are numerous attempts in the literature to make $f(g_i)\approx 0$ by introducing
more particles, like extra scalars or fermions \cite{Kundu:1994bs,grzadkowski,drozd,Chakraborty:2012rb,Chakraborty:2014oma,Chakraborty:2014xqa,Chakraborty:2016izr}.
While these attempts were more or less successful and provided some important constraints on the 
parameter space, the VC could hardly be stabilised over the entire energy scale from $v$ to $\Lambda$ 
if one considers the renormalisation group (RG) evolution of the couplings. This remains one of the major 
shortcomings of the bottom-up approach. 

We will take the bottom-up approach to its extreme limit. For us, whatever New Physics (NP) exists 
there at the high energy scale can be effectively integrated out at the scale $\Lambda$ to give us the SM, plus 
some effective operators involving only the SM fields, which is known as the SM Effective Field Theory 
(SMEFT). We will not venture to investigate the possible
nature of the UVC theory; rather, all the UVC information will be clubbed in the Wilson coefficients (WC)
of the effective operators. 

In SMEFT, the first interesting higher dimensional operators come at $d=6$ (the $d=5$ 
Weinberg operator is not relevant for scalar self-energies). There are many equivalent bases to 
express the complete set of $d=6$ operators. We will use the basis given in Ref.\ \cite{Corbett:2014ora}.   
Only a handful among the 59 dimension-six operators contribute to the quadratically divergent part of
the scalar self-energy. 

An $n$-dimensional operator can at most result in a divergence in Higgs self-energy
 that goes as $\Lambda^{n-2}$. As these operators are suppressed by $\Lambda^{n-4}$, one expects 
 contributions to $f(g_i)$ from all orders. What, then, is the rationale to consider only $d=4$ and $d=6$ operators? 
We have tried to answer this question in Section \ref{sec:dim8}. 

Thus, we will focus only on an effective theory with a schematic Lagrangian
\be
{\cal L} = c_{4i} O_i^{d=4} + \frac{1}{\Lambda^2} c_{6i} O_i^{d=6}\,,
\label{eq:dim-6lag}
\ee
where $c_{4i}$ and $c_{6i}$ are dimensionless constants. The VC now takes the form
\be
F \left(c_{4i}, c_{6i}\right) \approx 0\,.
\ee
Our aim will be to find out the parameter space for the $c_{6i}$ coefficients. 

In Section \ref{sec:dim8}, we discuss why it is enough to take into account only the dimension-six operators. 
In Sections \ref{sec:smvc} and \ref{sec:dim6}, we discuss the VC in the SM (with dimension-4 operators) and 
in SMEFT with dimension-six operators. In Section \ref{sec:result}, we show the allowed parameter space for the 
$c_{6i}$ coefficients and discuss our results. Section \ref{sec:conc} concludes the paper.

\section{Why we can neglect $d=8$ and higher operators}   \label{sec:dim8}

The $d=6$ SMEFT  has been well-explored, and there are several equivalent bases
to express all the $d=6$ operators. While the $d=8$ operators are not that well-investigated, it is known
\cite{Henning:2015alf} that there are 993 such operators with one generation and 44807 operators with three 
generations. A list of the relevant bosonic operators can be found in Ref.\ \cite{Hays:2018zze}. 

For $d=6$ operators, there are two types of diagrams that come with a $\Lambda^4$ divergence. First are 
the two-loop diagrams, like the one from $\left( \Phi^\dag\Phi\right)^3$, where $\Phi$ is the SM doublet 
Higgs field. The second class consists of one-loop diagrams but momentum-dependent vertices, like the 
one coming from $\left(D^\mu\Phi\right)^\dag\left(D_\mu\Phi\right) \Phi^\dag\Phi$. If the derivatives act 
on the internal scalar lines, the vertex has a momentum-dependence $\sim k^2$, where $k$ is the loop
momentum to be integrated over, and the resulting divergence 
is again quartic. However, there is a crucial difference: the first set comes with $(16\pi^2)^{-2}$, and the 
second set only with $(16\pi^2)^{-1}$, similar to the $d=4$ operators. Therefore, it is the one-loop diagrams that 
should be the most relevant in calculating the VC. One can have a similar conclusion with operators 
involving the gauge field tensors, and the final result is:

{\em Only those dimension-six operators contribute quartic divergences at one-loop for which both the derivatives 
act on the field in the loop.}

Thus, among the $d=8$ operators, one should look only for those operators that come with four derivatives, 
{\em i.e.}, $D^4$. There are only three such operators \cite{Hays:2018zze}, and all of them have a generic
structure of $(D\Phi)^\dag(D\Phi)(D\Phi)^\dag (D\Phi)$. As two of the derivatives act on the external leg fields
and hence give the square of the external leg momentum, the vertex factor can only have a $k^2$
dependence, and the divergence remains only $\Lambda^4$ and not $\Lambda^6$. 
Similarly, operators of the form $D^2(\Phi^2 W^2)$, where $W$ is the generic gauge tensor, do not generate 
any $\Lambda^6$ divergence. Thus, it is enough, 
within the limits of uncertainty, to consider only the $d=6$ operators. However, one may note that the 
argument is not watertight; the huge number of $d=8$ operators may offset the extra suppression coming
from $1/16\pi^2$. 

One can easily extend the logic for $d>8$ operators. For $d=2n$, one needs operators of the generic form 
$\Phi^\dag\Phi (D^pS)^\dag (D^pS)$, where $p=n-2$ and $S$ some scalar field, to throw a momentum 
dependence of $k^{2p}$ in the loop and make the one-loop diagram equally important as those coming from
$d=6$ operators. Apparently, all such operators can be reduced to harmless ({\em i.e.}, not producing 
a one-loop $\Lambda^4$ divergence) forms through equations of motion.   

\section{Veltman condition with dimension-4 operators} \label{sec:smvc}

We start from the SM Higgs potential with only $d\leq 4$ terms:
\be
V(\Phi)
= -\mu^2\Phi^\dag\Phi + \lambda (\Phi^\dag\Phi)^2 
\ee
where $\Phi$ is the SM doublet, with $\bra\Phi\ket = v/\sqrt{2}$. 
The Higgs self-energy receives a quadratically divergent correction
\be
\delta m_h^2 = \frac{\Lambda^2}{16\pi^2} \left(6\lambda + \frac34 g_1^2 + \frac94 g_2^2
- 6 g_t^2\right)\,,
\label{smvc}
\ee
where $g_1$ and $g_2$ are the $U(1)_Y$ and $SU(2)_L$ gauge couplings, and $g_t = \sqrt{2}
m_t/v$ is the top quark Yukawa coupling. All other fermions are treated as massless. 
Dimensional regularisation does not differentiate between quadratic and logarithmic divergences, and we get a
slightly different correction \cite{Einhorn:1992um}:
 \be
\delta m_h^2 \propto \frac{1}{\epsilon} \left(6\lambda + \frac14 g_1^2 + \frac34 g_2^2
- 6 g_t^2\right)\,.
\ee
As our goal is to cancel the strongest divergence, we will use the cut-off regularisation.
Two- and higher-loop diagrams can also contribute to the quadratic divergence, 
but they are suppressed from the one-loop contributions by a factor of 
$\ln(\Lambda/\mu)/16\pi^2$ or more, where $\mu$ is the regularisation scale. 
We will, therefore, not consider anything beyond one loop. 

At this point, let us make some comments on the gauge dependence of the VC. They are gauge independent, 
as can be explicitly checked by working out the quadratic divergences in Landau and 't~Hooft-Feynman gauge.
However, one may ask what happens in the unitary gauge, as the gauge propagator has a leading momentum 
dependence of $k^0$. While one may question the justification to use the unitary gauge as the VC is relevant only 
for $\Lambda \gg v$ where the electroweak symmetry is still unbroken, all particles are massless, and the condition
is formulated in terms of the couplings only, it would nevertheless be satisfactory to see that nothing 
catastrophic happens in the unitary gauge. For example, one may think of having a quartic divergence, $\sim 
\Lambda^4$, coming from the gauge loop, as the gauge propagator is not momentum suppressed. At the same
time, one has to remember that the electroweak symmetry is broken, and there are generic Higgs-gauge-gauge 
vertices in the theory. One can have a self-energy contribution with two such vertices, which again is 
quartically divergent. We have explicitly checked that these quartic divergences cancel; for the $W$-loop, the
amplitude with the four-point vertex gives $g^2 \Lambda^4/(128 \pi^2 m_W^2)$, which is exactly 
cancelled by the amplitude with two three-point vertices. The latter also gives a quadratic divergence, which
is needed to restore the gauge invariance.   

One can say that the quadratic divergence is under control if $|\delta m_h^2| \leq m_h^2$, 
which translates into\footnote{The VC can be expressed in terms of the masses only after the 
electroweak symmetry is broken.}
\be
\left\vert m_h^2 + 2m_W^2 + m_Z^2 - 4m_t^2\right\vert \leq \frac{16\pi^2}{3} \frac{v^2}{\Lambda^2} m_h^2\,.
\label{smvc2}
\ee
This inequality is clearly not satisfied in the SM for $v^2/\Lambda^2 \leq 0.1$, or $\Lambda \geq 760$ GeV,
and onset of NP at such a low scale is already ruled out by the LHC. Thus, one needs extra degrees of freedom,
like more scalars or fermions. There are a number of such studies in the literature; we refer the reader to,
{\em e.g.}, Refs.\ 
\cite{Kundu:1994bs,Chakraborty:2012rb,Chakraborty:2014oma,Chakraborty:2014xqa,Chakraborty:2016izr}.

\section{Veltman condition with dimension-six operators} \label{sec:dim6}
 
We will use the SMEFT basis as in Ref.\ \cite{Corbett:2014ora}. Keeping in mind that only operators with two or more
 Higgs fields are relevant and the divergence should be quartic, the relevant operators are as follows:

\begin{align}
& O_{WW} = \Phi^\dag \widehat{W_{\mu \nu}}\, \widehat{W^{\mu \nu}}\Phi\,,\ \ 
 O_{BB} = \Phi^\dag \widehat{B_{\mu \nu}}\, \widehat{B^{\mu \nu}}\Phi\,,\ \ 
O_{GG} = \Phi^\dag \Phi \widehat{G_{\mu \nu}}\, \widehat{G^{\mu \nu}}\,,\n\\
& O_W = (D_\mu \Phi)^\dag \widehat{W^{\mu \nu}} (D_\nu \Phi)\,,\ \
O_B = (D_\mu \Phi)^\dag \widehat{B^{\mu \nu}} (D_\nu \Phi)\,,\ \
O_{\phi,1} = (D_\mu \Phi)^\dag \Phi\, \Phi^\dag (D^\mu \Phi)\,,\n\\ 
& O_{\phi,2} = \frac12\, \partial ^\mu(\Phi^\dag \Phi)\,  \partial_\mu (\Phi^\dag \Phi)\,,\ \
O_{\phi,3} = \frac13\, (\Phi^\dag \Phi)^3\,,\ \
O_{\phi,4} = (D_\mu \Phi)^\dag (D^\mu \Phi)\, \Phi^\dag \Phi\,, 
\end{align}
where
\be
\widehat{B_{\mu \nu}} = \frac{ig'}{2}\, B_{\mu \nu}\,,\ \ 
\widehat{W_{\mu \nu}} = \frac{ig}{2}\, \sigma^a W^a_{\mu \nu}\,,\\
\widehat{G_{\mu \nu}} = \frac{ig_s}{2}\, \lambda^A G^A_{\mu \nu}\,,
\ee
$g$, $g'$ being the $SU(2)_L$ and $U(1)_Y$ gauge couplings respectively, and $\lambda^A$, $\sigma^a$ are the Gell-Mann and Pauli matrices. Note that the mixed gauge operator 
$O_{BW} = \Phi^\dag \widehat{B_{\mu \nu}}\, \widehat{W^{\mu \nu}}\Phi$ 
cannot generate a self-energy amplitude, either at one- or at two-loop. There is no contribution 
from $O_B$ either, due to the abelian nature of the field tensor, but we keep it for completeness.
Only the $O_{VV}$ ($V=W,B,G$) and $O_{\phi,i}$ ($i=1,2,4$) operators are relevant for one-loop diagrams.  

We would like to mention that the SMEFT basis is far from unique. The same exercise could have been performed
in the `Warsaw' basis \cite{Grzadkowski:2010es}, and the relevant operators fall in the $\varphi^6$,
$\varphi^4 D^2$, and $X^2 \varphi^2$ categories. For the relevant Feynman rules, we refer the reader to 
Ref.\ \cite{Dedes:2017zog}. 

\begin{figure}[htbp]
\centerline{
\includegraphics[width=12cm]{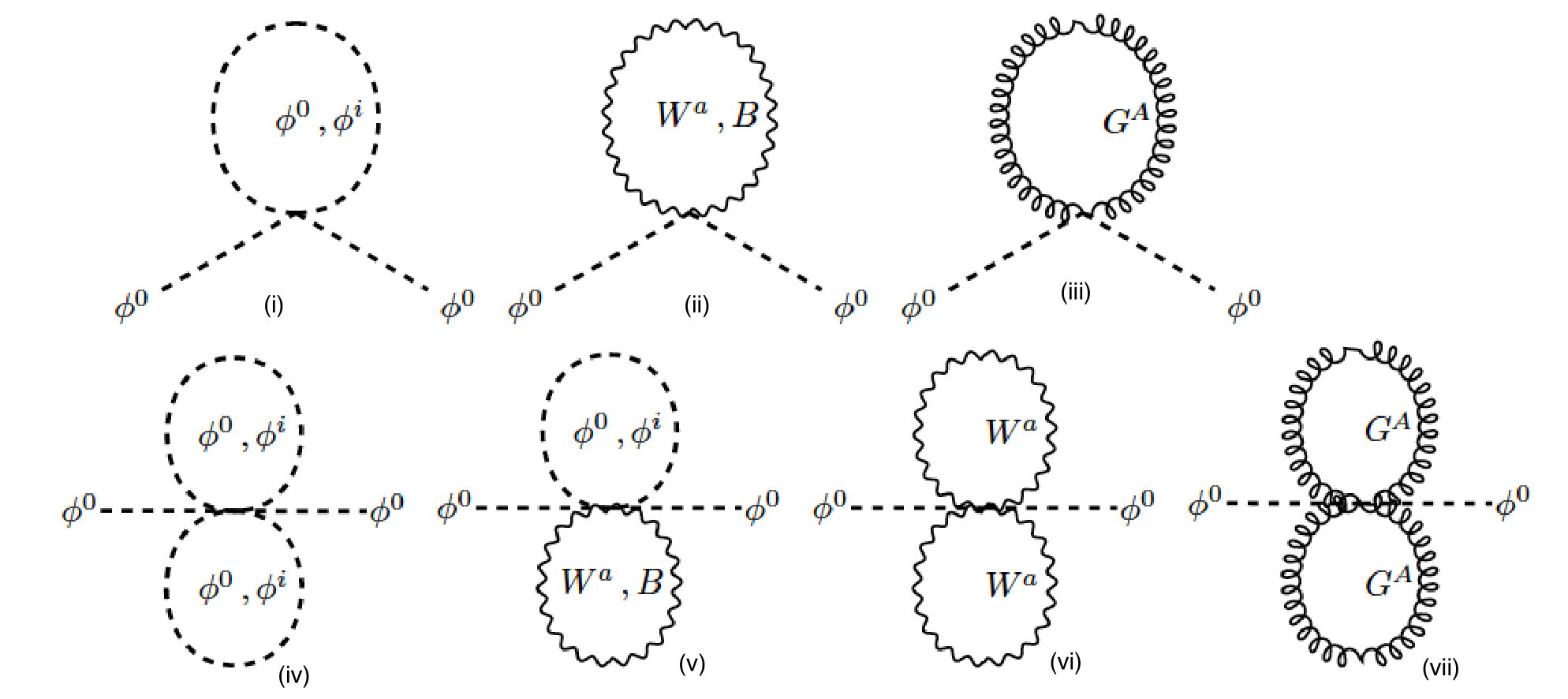}
}
\caption{The Feynman diagrams that contribute to the $\Lambda^4$ divergences. The first row shows the 
one-loop diagrams with momentum-dependent vertices; the second row shows the two-loop diagrams
where the vertex factor does not contain the loop momentum. The latter set is suppressed by an extra
$1/16\pi^2$ compared to the former set. $\phi^0$, $\phi^i$, $W^a$, $B$, and $G^A$ stand for the Higgs
boson, the Goldstone bosons, the SU(2)$_L$ and U(1)$_Y$ gauge bosons, and the gluons respectively.
The indices $i$ and $a$ run from 1 to 3, while $A$ runs from 1 to 8.}
\label{fig:feynman} 
\end{figure}

With these set of nine operators, we can write the dimension-six part of Eq.\ (\ref{eq:dim-6lag}) as
\be
{\cal L}_6 = \frac{1}{\Lambda^2} \sum_{i=1}^{9}\, c_{i} O_i\,,
\label{eq:dim-6lag2}
\ee
and Eq.\ (\ref{smvc}) takes the form
\be
\delta m_h^2 = \frac{\Lambda^2}{16\pi^2} \left(6\lambda + \frac34 g_1^2 + \frac94 g_2^2
- 6 g_t^2\right) + \frac{\Lambda^2}{16\pi^2} \sum_i f_i + \frac{\Lambda^2}{(16\pi^2)^2} \sum_i g_i\,,
\label{dim6vc1}
\ee
where $f_i$ and $g_i$ terms come respectively from the one-loop and two-loop quartic divergences
with the insertion of the operator $O_i$, any of the eight (except $O_B$) dimension-six operators listed above. 
The relevant Feynman diagrams are shown in Fig.\ \ref{fig:feynman}, where the top three are one-loop 
with momentum-dependent vertices contributing to $f_i$, and the bottom four, contributing to $g_i$, are 
two-loop where the vertex factors are momentum-independent; the latter set is suppressed by an 
extra $1/16\pi^2$.  In Appendix A, we show explicitly which operators contribute to which diagrams,
and why operators like $O_B$ are irrelevant for us. 

The contributions are given by
\bea
f_{\phi,1} =  -3\, c_{\phi,1}  \,, \ \ && \ \ g_{\phi,1} =  -\frac92\, \left({g'}^2+3g^2\right)\, c_{\phi,1}  \,,\n\\
f_{\phi,2} = -6\, c_{\phi,2} \,, \ \ && \ \ g_{\phi,2} = 0   \,,\n\\
f_{\phi,3} = 0  \,, \ \ && \ \ g_{\phi,3} =  18\, c_{\phi,3}  \,,\n\\
f_{\phi,4} = -3\, c_{\phi,4}  \,, \ \ && \ \ g_{\phi,4} =  -\frac92\, \left({g'}^2+3g^2\right)\, c_{\phi,4}  \,,\n\\
f_{WW} = -\frac{9}{4}\, g^2\, c_{WW}  \,, \ \ && \ \ g_{WW} = -27\, g^4\, c_{WW}   \,,\n\\
f_{BB} = -\frac34\, {g'}^2\, c_{BB}  \,, \ \ && \ \ g_{BB} =  0  \,,\n\\
f_{GG} = -6\, g_s^2\, c_{GG}  \,, \ \ && \ \ g_{GG} = -72\, g_s^4\, c_{GG}   \,,\n\\
f_{W} = 0  \,, \ \ && \ \ g_{W} = -\frac{27}{2}\, g^4\, c_W   \,,\n\\
f_{B} = 0 \,, \ \ && \ \ g_{B} = 0   \,.
\label{eq:f-g}
\eea
Eqs. (\ref{dim6vc1}) and (\ref{eq:f-g}) are the central results of this paper.  
The extra $1/16\pi^2$ suppression tells us that we may neglect the $g_i$ terms (and thus
will be justified to neglect the dimension-8 and other higher-dimensional operators), unless we deal with 
pathological cases like $\sum f_i \approx 0$, or all the WCs being zero except $c_{\phi,3}$.

With only the $f_i$ terms, the modified VC reads
\be
\frac{1}{16\pi^2}\left[
 \left(6\lambda + \frac34 g_1^2 + \frac94 g_2^2- 6 g_t^2\right) + \sum_i f_i \right] \leq \frac{\delta m_h^2}
 {\Lambda^2}\,,
\label{dim6vc2}
\ee
with all the couplings and WCs evaluated at the scale $\Lambda$. Eq.\ (\ref{dim6vc2}) immediately 
tells us that at least one, or perhaps more, WCs should be negative. As the operators do not
contain strongly interacting fields (except $O_{GG}$), the running between $\Lambda$, which is the matching
scale, and the electroweak scale, are controlled by electroweak radiative corrections only (at the leading order). 

Before we go into the next Section, let us enlist once more a couple of important points.

\begin{itemize}
\item If we assume the WCs at the matching scale are of order unity (which is expected if the UVC is 
perturbative in nature), we can safely neglect the two-loop $g_i$ terms, as they are suppressed by at least two 
orders of magnitude coming from $1/16\pi^2$. This also ensures that we do not have to worry about 
$d > 6$ operators. 

\item At what scale should the VC be satisfied? Obviously, it should be at the matching scale $\Lambda$. 
Eq.\ (\ref{smvc2}) shows that the cancellation need not be exact, it should be of the order of $v^2/\Lambda^2$. 
Thus, it is meaningless to talk about the fine-tuning problem if $\Lambda = 1$ TeV, and anyway we already know 
that there is no new physics (at least strongly interacting) at that scale, thanks to LHC. $\Lambda = 
100$ TeV makes the fine-tuning problem come back in a softened avatar, so this should be the correct ballpark
to study the issue. Even higher values, like $\Lambda = 10^6$ TeV, makes the fine-tuning problem seriously 
uncomfortable.

\end{itemize} 

\section{Result} \label{sec:result}
Following what has been said just now, we will study the VC for two values of $\Lambda$, namely, 100 TeV
and $10^6$ TeV. To start with, let us assume that only one of the six SMEFT operators 
(neglecting $O_{\phi,3}$, $O_W$, and $O_B$ which do not contribute to $f_i$) is present at the matching scale. 
We also need to evolve the SM couplings to that 
scale, for which we use the package {\tt SARAH} v4.14.1 \cite{Staub:2008uz}, with two-loop RG equations. 

Taking $\Lambda=100$ TeV, one gets, for exact cancellation of the quadratic divergence, 
\be
100~{\rm TeV} \ \ | \ \ c_{\phi,1} = c_{\phi,4} = 2c_{\phi,2}  = -1.15\,,\ \  c_{BB} =  -21.5\,, \ \ c_{WW} =  -4.13\,, \ \ 
c_{GG} = -0.78\,,
\label{eq:single100}
\ee
and for $\Lambda=10^6$ TeV
\be
10^6~{\rm TeV} \ \ | \ \ c_{\phi,1} = c_{\phi,4} = 2c_{\phi,2}  = -1.03\,,\ \  c_{BB} = -17.3\,, 
\ \ c_{WW} = -4.20\,, \ \ c_{GG} = -1.11\,.
\label{eq:single100}
\ee
Of course, one may relax these numbers a bit if exact cancellation is not warranted. Note the large 
values for the weak gauge WCs, they stem from the definition of the corresponding $f_i$s in Eq.\ (\ref{eq:f-g})
which contain $g^2$ or ${g'}^2$; the UVC need not be non-perturbative. 
On the other hand, if we take $\Lambda = 2$ TeV only (which is hardly of any practical interest), 
the corresponding exact-cancellation values are 
\be
2~{\rm TeV} \ \ | \ \ c_{\phi,1} = c_{\phi,4} = 2c_{\phi,2}  = -1.34\,,\ \  c_{BB} = -26.2\,, \ \ c_{WW} = -4.53\,, \ \ 
c_{GG} = -0.66\,.
\label{eq:single2}
\ee
This change is entirely due to the running of the SM couplings.

However, there is hardly any UVC theory that generates only one of these six operators at the matching scale. 
As the sign of the WCs can be either positive or negative, the six free parameters do not even give a closed 
hypersurface in the 6-dimensional plot, and therefore marginalisation is of very limited use.
Let us consider two distinct cases where only a pair of WCs are nonzero at $\Lambda$:\\
(1) Only $c_{\phi,2},c_{\phi,4} \not=0$: The approximate condition to satisfy the VC is
\bea
c_{\phi,4} + 2 c_{\phi,2} + 1.150 &=& 0 \ \ (\Lambda = 100~{\rm TeV})\,,\nonumber\\ 
c_{\phi,4} + 2 c_{\phi,2} + 1.030 &=& 0 \ \ (\Lambda = 10^6~{\rm TeV})\,.
\eea
(2) Only $c_{WW}, c_{BB} \not = 0$:
\bea
c_{BB} + 5.212 c_{WW} + 21.544 &=& 0  \ \ (\Lambda = 100~{\rm TeV})\,,\nonumber \\  
c_{BB} + 4.122 c_{WW} + 17.306 &=& 0  \ \ (\Lambda = 10^6~{\rm TeV})\,.
\eea 
All the WCs are evaluated at the scale $\Lambda$. The exact conditions broaden out to finite-width 
bands if we allow a finite amount of fine-tuning, the bands getting narrower for higher values of 
$\Lambda$.  

%

The SMEFT operators contribute to anomalous trilinear and quartic gauge-gauge and gauge-Higgs couplings,
as well as modified wavefunction renormalisation for the bosonic fields. It is indeed heartening to note that
the parameter space that we obtain is consistent with all other theoretical and experimental constraints
\cite{Almeida:2018cld,Almeida:2020ylr}. 
For other collider signatures of these $d=6$ operators, like vector boson scattering and 
Higgs pair production at the LHC, we refer the reader to, {\em e.g.}, Refs.\ \cite{Gomez-Ambrosio:2018pnl} and 
\cite{Liu-Sheng:2017pxk}.

\section{Conclusion}    \label{sec:conc}

In this paper, we have discussed the Veltman condition leading to the cancellation of the quadratic divergence
of the Higgs self-energy in the context of an SMEFT framework. In other words, we assume the existence of a cut-off 
scale $\Lambda$, below which we have the SM, while the theory above $\Lambda$ introduces higher-dimensional 
operators in the low-energy domain. If $\Lambda$ is large enough, the low-energy theory is still plagued 
by the $\sim \Lambda^2$ divergence, even if it is not as uncomfortable as what one gets with a desert up to
the Planck scale. 

We show that the higher dimensional operators lead to quadratic divergences too, but there are two distinct sources
of them. For example, with $d=6$ operators, such divergences can come from one-loop diagrams with 
momentum-dependent vertices, or two-loop diagrams with momentum-independent vertices. The latter, however, 
are suppressed by an extra loop factor of $1/16\pi^2$ and hence can be neglected as a first approximation. 
The same logic leads to the important point that only $d=6$ operators are relevant for such one-loop 
quadratic divergences. (There is a caveat, though: the number of relevant effective operators increases 
almost exponentially with $d$, and the loop suppression may just be compensated by the large number of 
such amplitudes.) 

We find that there are only six operators that contribute to the Veltman condition at the one-loop level. 
It turns out that at least one of the WCs has to be negative, but they are all consistent with a high-scale perturbative 
theory. The parameter space that we find is compatible with other theoretical and experimental constraints. Thus, this 
study should set a benchmark for the model builders.

\noindent {\em Acknowledgements} --- A.K.\ acknowledges the support from the Science and Engineering Research 
Board, Govt.\ of India, through the grants CRG/2019/000362, MTR/2019/000066, and DIA/2018/000003.

\appendix
\setcounter{equation}{0}
\renewcommand{\theequation}{\thesection.\arabic{equation}}
\section{The dimension-six operators and the Feynman diagrams}

Let us refer to Fig.\ \ref{fig:feynman}, which shows all the possible one- and two-loop diagrams originating 
from the set of dimension-six SMEFT operators. These are the only operators for which the divergence is 
quartic in nature. We will denote the first three one-loop diagrams by (i)-(iii) and the last four two-loop 
diagrams by (iv)-(vii) respectively. The diagrams corresponding to the amplitudes generated by different 
terms of any given operator are marked within square brackets. Note that the terms involving the gauge 
fields result in gauge-dependent propagators, but the final results are gauge independent.

\bea
O_{\phi,1} &=& (D_\mu \Phi)^{\dagger}\Phi \Phi^{\dagger}(D^\mu \Phi)\nonumber\\
&\supset& (\partial_\mu \Phi^{\dagger})\Phi\Phi^{\dagger}(\partial^\mu \Phi)  \ \ [{\rm i}]
+ \left[\frac{g^2}{4}\sigma^a \sigma^b \Phi^{\dagger} W_\mu ^{a }\Phi\Phi^{\dagger} W^{\mu b} \Phi 
+\frac{{g'}^2}{4}\Phi^{\dagger} B_\mu \Phi\Phi^{\dagger} B^{\mu}\Phi \right]  \ \ [{\rm v}]\,,\\
O_{\phi,2} &=& \frac{1}{2}\, \partial^\mu(\Phi^{\dagger}\Phi)\partial_\mu(\Phi^{\dagger}\Phi) \ \ [{\rm i}]\,,\\
O_{\phi,3} &=& \frac{1}{3}\, (\Phi^{\dagger}\Phi)^3 \ \ [{\rm iv}]\,,\\
O_{\phi,4} &=& (D_\mu \Phi)^\dag(D^\mu \Phi)(\Phi^{\dagger}\Phi)\nonumber\\
&\supset & (\partial_\mu \Phi^{\dagger})(\partial^\mu \Phi)\Phi^{\dagger}\Phi  \ \ [{\rm i}]
+ \left[ 
\frac{g^2}{4}\sigma^a \sigma^b \Phi^{\dagger} W_\mu ^{a } W^{\mu b} \Phi \Phi^{\dagger}\Phi +
\frac{{g'}^2}{4}\Phi^{\dagger} B_\mu  B^{\mu}\Phi \Phi^{\dagger}\Phi \right]  \ \ [{\rm v}]\,,\\
O_{BB} & = & \Phi^{\dagger}\widehat{B_{\mu \nu}} \widehat{B^{\mu \nu}} \Phi \nonumber\\
& \supset& -\frac{{g'}^2}{2}\Phi^\dagger\left[\partial_\mu B_\nu \partial^\mu B^\nu - 
\partial_\mu B_\nu \partial^\nu B^\mu\right] \Phi \ \ [{\rm ii}]\,,
\eea
\bea
O_{WW} & = & \Phi^{\dagger} \widehat{W_{\mu \nu}} \widehat{W^{\mu \nu}} \Phi  \nonumber\\
& \supset& \left[-\frac{g^2}{2} \, \Phi^\dagger [\sigma^a (\partial_\mu W^a_{\nu}) \sigma^p (\partial^\mu W^{p \nu})-\sigma^a (\partial_\mu W^a_{\nu}) \sigma^p (\partial^\nu W^{p \mu})]\Phi\right]
\ \ [{\rm ii}]\nonumber\\
&& -\frac{g^4}{4}\sigma^a f_{abc} \sigma^p f_{pqr}\Phi^\dagger W^b_{\mu} W^c_{\nu} W^{q \mu} W^{r \nu} \Phi
\ \ [{\rm vi}]\,,\\
O_{W} & = & (D_\mu \Phi)^\dagger \widehat{W^{\mu \nu}} (D_\nu \Phi) \nonumber\\
& \supset &
\left[- \frac{g^2}{4}  
(\partial_\mu \Phi)^\dagger \sigma^p [\partial^\mu W^{p \nu}-\partial^\nu W^{p \mu}]\sigma^b W^b_\nu \Phi 
+ \frac{g^2}{4}  \Phi^\dag W^a_\mu \sigma ^a \sigma ^p [\partial^\mu W^{p \nu}-\partial^\nu W^{p \mu} ]\partial_\nu \Phi \right]
\ \ [{\rm ii}] \nonumber\\
&& - \frac{ig^4}{8}\Phi^\dag {W^a_\mu} \sigma^a \sigma^p f_{pqr} W^{q \mu} W^{r \nu} \sigma^b W^b_\nu \Phi 
\ \ [{\rm vi}]\,, 
\\
O_{B} & = & (D_\mu \Phi)^\dag \widehat{B^{\mu \nu}} (D_\nu \Phi) \nonumber\\
& \supset & 
\left[\frac{{g'}^2}{4} \Phi^{\dagger} B_\mu[\partial^\mu B^\nu- \partial^\nu B^\mu] \partial_\nu \Phi -\frac{{g'}^2}{4}(\partial_\mu \Phi)^\dagger [\partial^\mu B^\nu-\partial^\nu B^\mu]  B_\nu \Phi \right]
\ \ [{\rm ii}]\,,\\
O_{GG} & = & \Phi^\dag \Phi \widehat{G_{\mu \nu}} \widehat{G^{\mu \nu}} \nonumber\\
& \supset &
\left[- \frac{g_s^2}{2} \,  
\Phi^\dag \left\{ \lambda^A \lambda^P \left(\partial_\mu G^A_{\nu}\, \partial^\mu G^{P \nu} -
\partial_\mu G^A_{\nu}\, \partial^\nu G^{P \mu}\right) \right\} \Phi \right]
\ \ [{\rm iii}]\nonumber\\
&&  - \frac{g_s^4}{4} \lambda^A \lambda^P f_{ABC} f_{PQR}\Phi^\dagger G^B_{\mu} 
G^C_{\nu} G^{Q \mu} G^{R \nu} \Phi \ \ [{\rm vii}]\,.
\end{eqnarray}
In the expressions above, lowercase and uppercase Latin stands for SU(2)$_L$ and SU(3)$_C$ indices, 
and run from 1 to 3, and 1 to 8, respectively. 

There are two comments that we would like to make here.
\begin{itemize}
\item 
Whether all the gauge terms contribute depend on the gauge choice; for example, the 
$\partial_\mu V_\nu \partial^\nu V^\mu$ term in $O_{VV}$ contributes in the 't Hooft-Feynman gauge, but not 
in the Landau gauge. As mentioned before, the final results are independent of the gauge choice. 

\item
The mixed-gauge operator $O_{BW}$ does not even come into the picture because there is no one- or two-loop
self-energy diagram originating from such an operator. On the other hand, $O_B$ can potentially generate a 
one-loop self-energy amplitude. However, note that one of the derivatives acts on the external leg, and so
the divergence can never be quartic. The same conclusion holds for $O_W$, but because of its non-abelian 
nature, there is a nontrivial two-loop amplitude. 

\end{itemize}


\end{document}